\documentclass[preprint,showpacs,preprintnumbers,amsmath,amssymb]{revtex4-1}
\usepackage{bm}
\usepackage{epsf}
\usepackage{color}

\everymath{\displaystyle}
\begin{document}           
\title{Exact form of the exponential Foldy-Wouthuysen transformation operator for an arbitrary-spin particle}

\author{Alexander J. Silenko}\email{alsilenko@mail.ru}
\affiliation{Research Institute for
Nuclear Problems, Belarusian State University, Minsk 220030,
Belarus\\
Bogoliubov Laboratory of Theoretical Physics, Joint Institute for
Nuclear Research, Dubna 141980, Russia}

\begin{abstract}
The exact form of the exponential Foldy-Wouthuysen transformation operator
applicable for a particle with an arbitrary spin is determined. It
can be successfully utilized for verifying any Foldy-Wouthuysen
transformation method based on the exponential operator. When a
verified method is relativistic, the relativistic exponential
operator should be expanded in the semirelativistic power series.
The obtained exponential operator can be also
used for a derivation of the Foldy-Wouthuysen Hamiltonian and its
comparison with Hamiltonians found by other methods. This
procedure makes it possible to check the validity of any other
method of the Foldy-Wouthuysen transformation.
\end{abstract}

 \maketitle


The Foldy-Wouthuysen (FW) transformation \cite{FW} restoring the
Schr\"{o}dinger form of relativistic wave equations is one of the
basic methods of contemporary quantum mechanics (QM). An
importance of the FW transformation for physics has significantly
increased nowadays due to the great progress of the art of analytic
computer calculations. A great advantage of the FW representation
is the simple form of operators corresponding to classical
observables. In this representation, the Hamiltonian and all
operators are even, i.e., block-diagonal (diagonal in two
spinors). The passage to the classical limit usually reduces to a
replacement of the operators in quantum-mechanical Hamiltonians
and equations of motion with the corresponding classical
quantities. The possibility of such a replacement, explicitly or
implicitly used in practically all works devoted to the FW
transformation, has been rigorously proved for the stationary case
in Ref. \cite{JINRLett12}. Thanks to these properties, the FW
representation provides the best possibility of obtaining a
meaningful classical limit of relativistic QM not only for the
stationary case \cite{FW,CMcK,JMPcond,JINRLett12} but also for the
nonstationary one \cite{PRAnonstat,nuclone}.

Various properties and applications of the FW method have been considered in Refs. \cite{BD,Urban,dVFor}.
The FW transformation is widely used in electrodynamics
\cite{electrodynamics,JMP}, quantum field theory
\cite{neznamovEChAYa}, optics \cite{Reuse,Khan,ultrafast},
condensed matter physics \cite{condensed}, nuclear physics
\cite{nuclone,nuclear}, gravity \cite{PRD,Gosgrav,ostgrav}, in the
theory of the weak interaction \cite{weak} and also in quantum
chemistry (see the books \cite{Dyall,ReiherWolfBook} and the
reviews
\cite{PengReiher,Autschbach,ReiherTCA,Liu,local,NakajimaH,ReiherArXivBook,ReiherRev}).
It is applicable not only for Dirac fermions but also for
particles with any spins
\cite{Case,Bryden,Pursey,Khan,Guertin,Leon,otherspins,PRDexact}.

The general form of an initial Hamiltonian for arbitrary-spin particles is given by \cite{PRA}
\begin{equation} {\cal H}=\beta{\cal M}+{\cal E}+{\cal
O},~~~\beta{\cal M}={\cal M}\beta, ~~~\beta{\cal E}={\cal E}\beta,
~~~\beta{\cal O}=-{\cal O}\beta. \label{eq3} \end{equation} The
even operators ${\cal M}$ and ${\cal E}$ and the odd operator
${\cal O}$ are diagonal and off-diagonal in two spinors,
respectively. Equation (\ref{eq3}) is applicable for a particle with any
spin if the number of components of a corresponding wave function
is equal to $2(2s+1)$, where $s$ is the spin quantum number. For a
Dirac particle, the ${\cal M}$ operator usually reduces to the
particle rest energy $mc^2$:
\begin{equation} {\cal H}_D=\beta mc^2+{\cal E}+{\cal
O}. \label{eq3Dirac} \end{equation}

The Hamiltonian ${\cal H}$ is Hermitian for fermions and pseudo-Hermitian (more exactly,
$\beta$-pseudo-Hermitian, ${\cal H}={\cal H}^\ddag\equiv\beta{\cal H}^\dag\beta$) for bosons. We assume that the
operators $\beta{\cal M},~{\cal E}$, and ${\cal O}$ also possess this property. The transformation operator for bosons is therefore $\beta$-pseudounitary
($U^\dag=\beta U^{-1}\beta$). We mention the existence of bosonic symmetries of the Dirac equation \cite{Simulik}.

The FW transformation operator can be presented in the two forms, nonexponential ($U_{FW}$) and exponential ($S_{FW}$):
\begin{equation} \Psi_{FW}=U_{FW}\Psi, \qquad U_{FW}=\exp{(iS_{FW})}.\label{Vvetott} \end{equation}

This operator transforms the initial Hamiltonian ${\cal H}$ to the FW representation:
\begin{equation} {\cal H}_{FW}=i\hbar\frac{\partial}{\partial
t}+U_{FW}\left({\cal H}-i\hbar\frac{\partial}{\partial
t}\right)U_{FW}^{-1}.
\label{taeq3}
\end{equation}

The FW Hamiltonian obtained with this operator is even. The FW transformation vanishes either lower or upper spinor for positive and
negative energy states, respectively.

There is an infinite set of representations different from the FW
representation whose distinctive feature is a block-diagonal
form of the Hamiltonian. The FW transformation is \emph{uniquely} defined by the condition that the exponential
operator $S_{FW}$ is \emph{odd},
\begin{equation} \beta S_{FW}=-S_{FW}\beta,
\label{VveEfrt} \end{equation} and Hermitian \cite{E,erik}
($\beta$-pseudo-Hermitian for bosons). This condition is equivalent to \cite{E,erik}
\begin{equation} \beta U_{FW}=U^\dag_{FW}\beta.\label{Erikcon} \end{equation}

Eriksen \cite{E} has found the exact expression for the nonexponential FW transformation operator. It is convenient to present this expression in the form \cite{JMPcond}
 \begin{equation}
U_{E}=U_{FW}=\frac{1+\beta\lambda}{\sqrt{2+\beta\lambda+\lambda\beta}},
~~~ \lambda=\frac{{\cal H}}{({\cal H}^2)^{1/2}}. \label{eqXXI}
\end{equation}
The initial Hamiltonian operator ${\cal H}$ is arbitrary. It
is easy to see that \cite{E}
\begin{equation}\lambda^2=1, \quad [\beta\lambda,\lambda\beta]=0, \quad [\beta,(\beta\lambda+\lambda\beta)]=0,\label{eq3X3}
\end{equation} where $[\dots,\dots]$ means a commutator.

The equivalent form of the operator $U_{E}$ \cite{JMPcond} shows that it is properly unitary ($\beta$-pseudounitary for bosons):
\begin{equation}
U_{E}=\frac{1+\beta\lambda}{\sqrt{(1+\beta\lambda)^\dag(1+\beta\lambda)}}.
\label{JMP2009}
\end{equation}

The Eriksen formula is an important achievement of the theory of the FW transformation.
However, the FW transformation method proposed by Eriksen \cite{E} is semirelativistic.
 We use the term ``semirelativistic'' for methods applying an expansion of a derived block-diagonal Hamiltonian in a series of even
terms of ascending order in $1/c$. For the semirelativistic and the relativistic methods, the zeroth order Hamiltonian is
the Schr\"{o}dinger one and the relativistic FW Hamiltonian of a free particle,
respectively. The Eriksen method
is not practically used in
specific calculations. Since the exact
equation (\ref{eqXXI}) contains the square roots
of Dirac matrices, it excludes a possibility to obtain a series
of relativistic terms with the relativistic FW Hamiltonian of a
free particle \cite{FW} as the zero-order approximation (see Ref. \cite{PRA2016} for more details). Contemporary methods of the FW transformation are relativistic. The use of all semirelativistic methods is restricted due to their divergence at large momenta, when $p/(mc)>1$. In this case (which takes place
for an electron near a nucleus), the semirelativistic methods
become inapplicable \cite{ReiherWolf} (see also Ref. \cite{PRA2016}).

Nevertheless, the calculation of the FW
Hamiltonian by the 
perfectly substantiated Eriksen method 
\cite{VJ,TMPFW,PRA2015} is very
important for checking results obtained by other semirelativistic and relativistic methods (see the examples given in Ref. \cite{PRA2016}).

We should mention that the condition (\ref{VveEfrt}) is used much more often than the condition (\ref{Erikcon}) to check the correspondence of the final Hamiltonian to the FW representation. For example, this is one of thebasic conditions for the generalized Douglas-Kroll-Hess 
method which is rather widely applied in quantum chemistry \cite{ReiherWolf,ReiherWolfNext,Highorder,ReiherRev}. This fact obviously demonstrates the importance of the exact derivation of the exponential FW transformation operator.
However, the problem of the exact form of $S_{FW}$ was never investigated. In the present study, we solve this problem.

Evidently, the exact form of $S_{FW}$ should be based on the operator $\lambda$. It is convenient to use the following relations (see, e.g., Ref. \cite{PRA2015}):
\begin{equation}\begin{array}{c}
{S}_{FW}=-i\beta\Theta, \qquad
U_{FW}=\cos{{S}_{FW}}+i\sin{{S}_{FW}}=\cos{\Theta}+\beta\sin{\Theta}. \end{array} \label{exptrol}
\end{equation}
To obtain an explicit expression for $S_{FW}$, it is instructive to consider the special case $[{\cal O},{\cal M}]=[{\cal O},{\cal E}]=[{\cal M},{\cal E}]=0$. When the operators ${\cal M},~{\cal E}$, and ${\cal O}$ do not explicitly depend on time and
\begin{equation}\begin{array}{c}
\tan {2\Theta}=\frac{{\cal O}}{{\cal M}}, \end{array} \label{exctndn}
\end{equation} the odd terms are eliminated and the Hamiltonian ${\cal H}$ is transformed to the FW representation \cite{PRA2015}. In the considered special case,
$$ \begin{array}{c} \left({\cal H}^2\right)^{1/2}=\epsilon+\frac{(\beta {\cal M}+{\cal O}){\cal E}}{\epsilon},\\
{\cal H}=(\beta {\cal M}+{\cal O})\left[1+\frac{(\beta {\cal M}+{\cal O}){\cal E}}{\epsilon^2}\right],\\ \epsilon=\sqrt{{\cal M}^2+{\cal O}^2},
\end{array} $$
and the operator $\lambda$ is determined exactly \cite{Valid}:
\begin{equation}
\lambda=\frac{\beta {\cal M}+{\cal O}}{\epsilon}.
\label{znak} \end{equation}

Equation (\ref{exctndn}) has two solutions, $\Theta_1$ and $\Theta_2$, differing in $\pi/2$ \cite{JMP}.
Since
$$\tan {2\Theta}=\frac{2\tan\Theta}{1-\tan^2\Theta},~~~\tan\Theta=
\frac{\tan {2\Theta}}{1\pm\sqrt{1+\tan^2 {2\Theta}}},   $$
they are defined by the relations
\begin{equation}
\tan \Theta_1=\frac {{\cal O}}{\epsilon+{\cal M}},~~~
\tan \Theta_2=-\frac {{\cal O}}{\epsilon-{\cal M}}.
\label{eq16}
\end{equation} As
\begin{equation}
\cos {2\Theta}=\frac{1-\tan^2\Theta}{1+\tan^2\Theta},~~~\sin
{2\Theta}=\tan {2\Theta}\cos {2\Theta}, \label{eqLL}\end{equation}
the needed trigonometrical operators are given by
\begin{equation}
\cos {2\Theta_1}=\frac{{\cal M}}{\epsilon},~~~\cos
{2\Theta_2}=-\frac{{\cal M}}{\epsilon},~~~\sin
{2\Theta_1}=\frac{{\cal O}}{\epsilon},~~~\sin
{2\Theta_2}=-\frac{{\cal O}}{\epsilon}. \label{eqMM}\end{equation}
This equation shows that $\cos {2\Theta_1}>0,~\cos {2\Theta_2}<0$.

Thus, there are two unitary transformations of the operator
${\cal H}$ to an even form. They are characterized by the
angles $\Theta_1$ and $\Theta_2$, where the angle $\Theta_1$ corresponds to the
FW transformation.
As a result of the both transformations, one of the spinors (lower
for $\Theta_1$ and upper for $\Theta_2$) becomes zero as for free
particles \cite{JMP}. Since we consider the FW transformation,
$\Theta=\Theta_1$.

A comparison with Eq. (\ref{exctndn}) shows that
\begin{equation}\begin{array}{c}
\sin {2\Theta}=\frac{1}{2}\left(\lambda-\beta\lambda\beta\right). \end{array} \label{exctlmd}
\end{equation} The corresponding exponential FW transformation operator is given by
\begin{equation}\begin{array}{c}
{S}_{FW}=-\frac{i\beta}{2}\arcsin{\frac{\lambda-\beta\lambda\beta}{2}}. \end{array} \label{expteot}
\end{equation}

Let us check that Eqs. (\ref{exctlmd}) and (\ref{expteot}) obtained in the above-mentioned special case remain valid in the general case of a relativistic particle in arbitrary stationary fields. We can utilize the operator relations (cf. Ref. \cite{PRA2015})
\begin{equation}\begin{array}{c}
\sin{\Theta} =\frac{\sin
{2\Theta}}{\sqrt{2\left(1+\cos{2\Theta}\right)}} =\frac{\sin
{2\Theta}}{\sqrt{2\left(1+\sqrt{1-\sin^2{2\Theta}}\right)}},
\end{array} \label{oprel}
\end{equation}
\begin{equation}\begin{array}{c}
\cos{\Theta} =\frac{1+\cos{2\Theta}}
{\sqrt{2\left(1+\cos{2\Theta}\right)}} 
=\frac{1+\sqrt{1-\sin^2{2\Theta}}}
{\sqrt{2\left(1+\sqrt{1-\sin^2{2\Theta}}\right)}}.
\end{array} \label{opren}
\end{equation}
These relations repeat the corresponding trigonometrical ones.
We can mention that the FW transformation
operator $U_{FW}$ is equal to unit when the initial Hamiltonian
does not contain odd terms (${\cal O}=0$).

As follows from Eq. (\ref{exptrol}), the
nonexponential FW transformation operator reads
\begin{equation}\begin{array}{c} U_{FW}=\frac{1+\sqrt{1-\sin^2{2\Theta}}+\beta\sin {2\Theta}} {\sqrt{2\left(1+\sqrt{1-\sin^2{2\Theta}}\right)}}. \end{array} \label{opnel}
\end{equation}

Equations (\ref{eq3X3}) and (\ref{exctlmd}) result in
\begin{equation}
1-\sin^2{2\Theta}=\frac14\left(\beta\lambda+\lambda\beta\right)^2.
\label{main}
\end{equation}

The use of Eqs. (\ref{exctlmd}) and (\ref{main}) shows the equivalence of Eqs. (\ref{eqXXI}) and (\ref{opnel}). This  equivalence rigorously proves the validity of Eq. (\ref{expteot}) in the general case.

Equation (\ref{expteot}) cannot be used for a derivation of the FW Hamiltonian in the \emph{relativistic} case because the operator $S_{FW}$ contains the square root of Dirac matrices (or corresponding matrices for particles with other spins). However, this equation opens a wonderful possibility to check a validity of semirelativistic and relativistic methods of the FW transformation. Indeed, a calculation of a \emph{semirelativistic} series for the exponential operator is straightforward.

The numerator and denominator in the formula for $\lambda$ commute. Therefore, it is convenient to present this quantity in the form
\begin{equation}
\lambda=\frac12\left\{{\cal H},\left({\cal H}^2\right)^{-1/2}\right\},
\label{mainlambda}
\end{equation} where $\{\dots,\dots\}$ means an anticommutator.
Then, we apply the expansion of the square root in the power series. It is important that this procedure can be performed for arbitrary-spin particles because the initial Hamiltonians can be presented in the form (\ref{eq3Dirac}). For example, the operator ${\cal M}$ for a spin-1 particle in a magnetic field reads \cite{TMPFW,PRDexact}
\begin{equation}{\cal M}=mc^2+\frac{\bm\pi^2}{2m}-\frac{e\hbar}{mc}\bm
S\cdot\bm B,\label{eqM}
\end{equation} where $\bm S$ is the spin matrix for the spin-1 particle, $\bm\pi=\bm p-(e/c)\bm A$ is the kinetic momentum operator, $\bm A$ is the vector potential, and $e$ is the charge of a particle. For the electron, it is negative ($e=-|e|$).
The two last terms in Eq. (\ref{eqM}) can be added to the operator ${\cal E}$ and therefore the operator ${\cal M}$ can be reduced to $mc^2$.

For analytic calculations with computer, one can use the formula
\begin{equation}\sqrt{{\cal H}^2}=mc^2\sqrt{1+\frac{{\cal H}^2-m^2c^4}{m^2c^4}}=
mc^2\sqrt{1+\frac{2\beta mc^2{\cal E}+{\cal O}^2+{\cal E}^2+
\{{\cal O},{\cal E}\}}{m^2c^4}} \label{forkorn}
\end{equation} and the well-known expansions
\begin{equation}\begin{array}{c} (1+x)^{-1/2}=1-\frac12x+\frac{1\cdot3}{2\cdot4}x^2-\frac{1\cdot3\cdot5}{2\cdot4\cdot6}x^3+\dots, \\
\arcsin{x}=x+\frac{1}{2\cdot3}x^3+\frac{1\cdot3}{2\cdot4\cdot5}x^5+\frac{1\cdot3\cdot5}{2\cdot4\cdot6\cdot7}x^7+\dots \end{array}\label{expansions}
\end{equation}

The result of the expansion in the power series can be written as follows:
\begin{equation}
\left({\cal H}^2\right)^{-1/2}=\frac{1+q_{\cal E}+q_{\cal
O}}{mc^2}, ~~~\beta q_{\cal E}=q_{\cal E}\beta,
~~~\beta q_{\cal O}=-q_{\cal O}\beta,
\label{mainexp}
\end{equation} where $q_{\cal E}$ and $q_{\cal O}$ denote the sums of even and odd terms, respectively. The resulting expansion of the operator $(\lambda-\beta\lambda\beta)/2$
in the semirelativistic power series is given by
\begin{equation}
\frac{\lambda-\beta\lambda\beta}{2}=\frac{1}{2mc^2}\bigl[2{\cal O}+\{{\cal E},q_{\cal
O}\}+\{{\cal O},q_{\cal
E}\}\bigr].
\label{lambdminus}
\end{equation}

The final equation for the exponential FW transformation operator takes the form
\begin{equation}\begin{array}{c}
{S}_{FW}=-\frac{i\beta}{2}\arcsin{\left(\frac{1}{2mc^2}\bigl[2{\cal O}+\{{\cal E},q_{\cal
O}\}+\{{\cal O},q_{\cal
E}\}\bigr]\right)}. \end{array} \label{expfinl}
\end{equation}
This operator can be calculated with any necessary precision while the computational effort 
depends on this precision.

Equation (\ref{expfinl}) gives one a wonderful opportunity to
verify any FW transformation method based on the exponential
operator. If the checked FW transformation method is relativistic,
one needs to expand all terms of the relativistic series for
$S_{FW}$ in powers of ${\cal E}/(mc^2),~{\cal O}/(mc^2)$.
Comparison with an explicit form of Eq. (\ref{expfinl}) verifies
the checked method. Of course, any expansion can be performed only
if the semirelativistic series is convergent (see Refs.
\cite{ReiherWolf,PRA2016}).

We should also add that the FW Hamiltonian can be calculated with
the exponential operator as follows (cf. Ref. \cite{FW}):
\begin{equation}  \begin{array}{c}
{\cal H}_{FW}={\cal H}+i[S_{FW},{\cal G}]+\frac{i^2}{2!}[S_{FW},[S_{FW},{\cal
G}]]+\frac{i^3}{3!} [S_{FW},[S_{FW},[S_{FW},{\cal G}]]]+\cdots, \end{array} \label{eqNEX} \end{equation}
where
\begin{equation}
{\cal G}={\cal H}-i\hbar\frac{\partial}{\partial t}, \qquad \left[S_{FW},\frac{\partial}{\partial t}\right]\equiv-\frac{\partial S_{FW}}{\partial t}.
 \label{eqXXVII} \end{equation}
Unlike the original FW approach \cite{FW}, this approach needs not subsequent iterations.

This result allows one to verify any other FW transformation
method which is not based on the exponential operator. For this
purpose, one has to expand all terms of the relativistic series
for the FW Hamiltonian calculated by the
checked method in powers of ${\cal E}/(mc^2),~{\cal
O}/(mc^2)$ and then has to compare the
obtained expression with the FW Hamiltonian given by Eqs.
(\ref{eqNEX}) and (\ref{eqXXVII}). The computational
effort, of course, grows when the maximum powers increase.
However, any error usually manifests itself in noncoincidence of the 
first few terms (see, for example, Refs. \cite{PRA2016,LiuPeng}).

Let us verify the original method by Foldy and Wouthuysen \cite{FW} as an example of the application of Eqs. (\ref{forkorn}) -- (\ref{expfinl}).
A calculation of the exponential FW transformation operator in the stationary case with allowance for all terms up to the order of $m^{-4}$ results in
\begin{equation}\begin{array}{c}
S_{FW}=-\frac{i}{2mc^2}\beta{\cal O}-\frac{i}{4m^2c^4}[{\cal O},{\cal E}]+\frac{i}{6m^3c^6}\beta{\cal O}^3\\
-\frac{i}{8m^3c^6}\beta[[{\cal O},
{\cal E}],{\cal E}]+\frac{3i}{16m^4c^8}\{{\cal O}^2,[{\cal O},
{\cal E}]\}-\frac{i}{16m^4c^8}\bigl[[[{\cal O},
{\cal E}],{\cal E}],{\cal E}\bigr].
\end{array} \label{Pro2p}\end{equation}

The original method \cite{FW} belongs to iterative methods. The result of successive
iterations expressed by the equation
\begin{equation} U=\ldots \exp{(iS^{(n)})}\ldots \exp{(iS''')}\exp{(iS'')}\exp{(iS')}\exp{(iS)} \label{Vvetgen} \end{equation}
can be presented in the exponential form with the use of the
Baker-Campbell-Hausdorff formula (see Ref. \cite{PRA2016} and references therein). This formula 
defines the product of two exponential operators:
\begin{eqnarray}
\exp(A)\exp(B)=\exp{\biggl(A+B+\frac12[A,B]+\frac{1}{12}[A,[A,B]]-\frac{1}{12}[B,[A,B]]}\nonumber\\ -\frac{1}{24}\bigl[A,[B,[A,B]]\bigr]+{\rm higher~ order~
commutators}\biggr).\label{EriKorl}\end{eqnarray}

To verify the original method \cite{FW}, it is sufficient to
take into account three first iterations and to hold terms up to
the order of $m^{-3}$. In the considered stationary case, they are
given by \cite{BD,dVFor,PRA2016}
\begin{equation}\begin{array}{c}
S=-\frac{i}{2mc^2}\beta{\cal O}, \qquad
S'=-\frac{i}{4m^2c^4}[{\cal O},{\cal E}]+\frac{i}{6m^3c^6}\beta{\cal O}^3,\\
S''=-\frac{i\beta}{8m^3c^6}[[{\cal O},
{\cal E}],{\cal E}].
\end{array} \label{Pro1p} \end{equation}

The result of these iterations can be approximately presented as follows (see Ref. \cite{PRA2016}):
\begin{equation}\begin{array}{c}
U=\exp{(iS'')}\exp{(iS')}\exp{(iS)}=\exp{(i\mathfrak{S})},\\ \mathfrak{S}=S''+S'+S-\frac{i}{2}[S,S']=-\frac{i}{2mc^2}\beta{\cal O}-\frac{i}{4m^2c^4}[{\cal O},{\cal E}]+\frac{i}{6m^3c^6}\beta{\cal O}^3\\
-\frac{i}{8m^3c^6}\beta[[{\cal O},
{\cal E}],{\cal E}]+\frac{i}{16m^3c^6}\beta[{\cal O}^2,{\cal E}].
\end{array} \label{Vvetgnm} \end{equation}

The operator $\mathfrak{S}$ differs from $S_{FW}$ due to the even last term. Therefore, the original method of the FW transformation \cite{FW} does not lead to the FW representation. This paradoxical fact has been first mentioned by Eriksen and Korlsrud \cite{erik} (see also Refs. \cite{dVFor,PRA2016,VJ,TMPFW}). The original method \cite{FW} can be corrected due to an additional transformation \cite{PRA2016}. This correction allows one to reach the FW representation.

Any representation which block-diagonalizes the Hamiltonian but
differs from the FW representation can be successfully used for a
calculation of energy spectrum of particles in stationary states.
However, an application of such a representation for other
purposes is restricted. In particular, the representation which
differs from the FW one is, at least, inconvenient for description
of spin processes (see, for example, Refs.
\cite{JMPcond,PRD,Quach2015}).

Thus, we have derived the exact exponential FW transformation
operator which is applicable for a particle with an arbitrary
spin. This result provides an opportunity to verify any FW
transformation method based on the exponential operator. Moreover,
the validity of any other FW transformation method can also be
checked. For this purpose, one has to use Eq.
(\ref{eqNEX}) in order to calculate the FW Hamiltonian and to
compare it with the corresponding Hamiltonian found by the checked
method. The latter possibility may need a greater computational
effort.

\section*{Acknowledgements}

This work was supported in part by the Belarusian Republican
Foundation for Fundamental Research, grant No. $\Phi$16D-004 and
by the Heisenberg-Landau program of the German Ministry for
Science and Technology (Bundesministerium f\"{u}r Bildung und
Forschung).

\end{document}